\documentclass[a4paper,11pt,twoside]{article}
\usepackage{amsmath}
\usepackage{multicol} 
\usepackage{amssymb}
\usepackage{floatflt}
\usepackage{graphicx}
\newlength{\myem}
\newcommand{\sep}[1]{#1}
\newcounter{mysubequation}[equation]
\renewcommand{\themysubequation}{\alph{mysubequation}}
\newcommand{\mytag}{\stepcounter{mysubequation}%
\tag{\theequation\protect\sep{\themysubequation}}}
\newcommand{\globallabel}[1]{\refstepcounter{equation}\label{#1}}

\newcommand{\GeV}{\,\mathrm{GeV}}

\newcommand{\ecm}{e\,\mathrm{cm}}

\DeclareMathOperator{\im}{Im}

\DeclareMathOperator{\diag}{diag}

\newcommand{\cu}{\mathscr{U}}
\newcommand{\A}{{\tilde A}}
\newcommand{\pdagger}{{\phantom{\dagger}}}
\newcommand{\gfun}{g}
\newcommand{\com}[1]{\Blue{\{\footnotesize\bf #1\}}\Black}
\renewcommand{\com}[1]{{}}

\def\CPV{{\begin{picture}(12,0)(0,0)
  \put(0,0){\scriptsize CP}\put(0,0){\line(2,1){12}}\end{picture}}}
\newcommand{\MT}{M_{\CPV}}

\newcommand{\be}{\begin{equation}}
\newcommand{\ee}{\end{equation}}
\newcommand{\ba}{\begin{array}}
\newcommand{\ea}{\end{array}}

\def\Red  {}

\def\Black{}
\def\Blue {}
\newcommand{\eq}[1]{~(\ref{eq:#1})}
\newcommand{\NP}{Nucl. Phys.}
\newcommand{\PRL}{Phys. Rev. Lett.}

\newcommand{\PL}{Phys. Lett.}
\newcommand{\PR}{Phys. Rev.}
\newcommand{\mb}[1]{\mbox{\normalsize\boldmath $#1$}}
\newcommand{\fig}[1]{~{\rm \ref{fig:#1}}}

\newcommand{\MGUT}{M_{\rm GUT}}

\font\tenrsfs=rsfs10
\font\sevenrsfs=rsfs7
\font\fiversfs=rsfs5
\newfam\rsfsfam
\textfont\rsfsfam=\tenrsfs
\scriptfont\rsfsfam=\sevenrsfs
\scriptscriptfont\rsfsfam=\fiversfs
\def\mathscr#1{{\fam\rsfsfam\relax#1}}

\def\circa#1{\,\raise.3ex\hbox{$#1$\kern-.75em\lower1ex\hbox{$\sim$}}\,}
\makeatletter
\def\art{\@ifnextchar[{\eart}{\oart}}
\def\eart[#1]#2#3#4#5#6{{\rm #2}, {\em #3 \bf #4} {\rm (#6) #5} ({\em #1})}
\def\hepart[#1]#2{{\rm #2, \em#1}}
\newcommand{\oart}[5]{{\rm #1}, {\em #2 \bf #3} {\rm (#5) #4}}

\newcounter{alphaequation}[equation]
\def\thealphaequation{\theequation\hbox to
0.6em{\hfil\alph{alphaequation}\hfil}}
\def\eqnsystem#1{
\def\@eqnnum{{\rm (\thealphaequation)}}
\def\@@eqncr{\let\@tempa\relax \ifcase\@eqcnt \def\@tempa{& & &} \or
  \def\@tempa{& &}\or \def\@tempa{&}\fi\@tempa
  \if@eqnsw\@eqnnum\refstepcounter{alphaequation}\fi
\global\@eqnswtrue\global\@eqcnt=0\cr}
\refstepcounter{equation} \let\@currentlabel\theequation \def\@tempb{#1}
\ifx\@tempb\empty\else\label{#1}\fi
\refstepcounter{alphaequation}
\let\@currentlabel\thealphaequation
\global\@eqnswtrue\global\@eqcnt=0 \tabskip\@centering\let\\=\@eqncr
$$\halign to \displaywidth\bgroup \@eqnsel\hskip\@centering
$\displaystyle\tabskip\z@{##}$&\global\@eqcnt\@ne
\hskip2\arraycolsep\hfil${##}$\hfil& \global\@eqcnt\tw@\hskip2\arraycolsep
$\displaystyle\tabskip\z@{##}$\hfil
\tabskip\@centering&\llap{##}\tabskip\z@\cr}
\def\endeqnsystem{\@@eqncr\egroup$$\global\@ignoretrue} \makeatother

\newcommand{\FNAL}{Fermi National Accelerator Laboratory, P.O.\ Box
500, Batavia, IL 60510, USA}

\newcommand{\preprintdate}{FermiLab--Pub--01/249-T}
\newcommand{\preprintnumber}{%
CERN--TH/2001--221\\
IFUP--TH/2001--23}
\newcommand{\hepnumber}{hep-ph/0108275} 
\newcommand{\titletext}{\Red\LARGE Electron and muon electric dipoles\\
in supersymmetric scenarios 
\Black} 
\newcommand{\authortext}{\vspace{-5mm}\large{\bf Andrea Romanino}$^{\, a}$ 
{\bf and Alessandro Strumia}$^{\, b}$\thanks{On leave from
dipartimento di Fisica dell'Universit\`a di Pisa and INFN.}
\medskip\\\em\normalsize $\mbox{}^a$ \FNAL \\[0.1\baselineskip]
$\mbox{}^b$ Theoretical physics division, CERN, CH-1211, Gen\`eve 23,
Suisse} 
\newcommand{\abstracttext}{
We study if a sizeable muon electric dipole can arise in supersymmetric frameworks able to account for the
tight experimental bounds on sfermion masses, like an appropriate flavor symmetry,
or like a flavor-blind mechanism of SUSY breaking
(in presence of radiative corrections charchteristic of GUT models,
or due to Yukawa couplings of neutrinos in see-saw models).
In some cases it is possible to evade the na\"{\i}ve scaling 
$d_\mu/d_e = m_\mu/m_e$ and obtain a $d_\mu$ as large as  $10^{-22\div 23} \ecm$.
In most cases $d_\mu$ is around $10^{-24\div 25}\ecm$ and
$(d_\mu/d_e)/(m_\mu/m_e)$
is only slightly different from one: this ratio contains interesting
informations on the source of the dipoles and on the texture of the
lepton Yukawa matrix.  We also update GUT predictions for $\mu\to e
\gamma$ and related processes.
}

\title{
\normalsize
\begin{tabular}[t]{l}\hepnumber\\\preprintdate\end{tabular}
\hspace*{\fill}
\begin{tabular}[t]{l}\preprintnumber\end{tabular}
\vspace{3\baselineskip}\\\Large\bfseries\titletext\bigskip}
\author{\begin{minipage}[t]{0.8\textwidth}
\normalsize\centering\authortext
\end{minipage}}
\date{}

\begin{document}
 \maketitle\Blue
\begin{abstract}\normalsize\noindent\large 
\abstracttext 
\end{abstract}\Black\normalsize\vspace{\baselineskip}

\section{Introduction}

The electric dipole moments (EDMs) of elementary particles represent a
powerful probe of physics beyond the Standard Model (SM). In the SM,
once the strong CP problem has been taken care of, the EDMs are far
beyond reach of foreseeable experiments, whereas supersymmetric
extensions of the SM may provide predictions in the range of
interest. Whereas the experimental~\cite{dNexp,deexp,dHgexp,dmuexp}
and theoretical~\cite{dipoliSoft,HKR,BH,DH,LFV} interest have mainly
focussed on the EDMs of the electron and light quarks (through neutron
and atomic EDMs), the recent prospects of improving the sensitivity to
the muon EDM $d_\mu$ by 5 orders of magnitude down to $d_\mu \sim
10^{-24}\ecm$~\cite{dmu} make $d_\mu$ an additional observable of
interest. 

We focus our attention on supersymmetric extensions of the SM, that
remain the relatively more promising solution of the Higgs mass
hierarchy `problem'~\cite{LEPparadox}, although no sparticles have been found at LEP.
The MSSM (i.e.\ the softly broken supersymmetric extension of the SM
with minimal field content and unbroken $R$-parity) potentially
contains additional sources of CP-violation associated to the
supersymmetry breaking part of the Lagrangian.  With no assumptions on
it, it is of course possible to find a region in the huge parameter space
that gives a detectable $d_\mu$. A complex muon $A$-term, $A_\mu$, is
the simplest possibility.  Ref.~\cite{FMS} discusses additional
possibilities. 

On the other hand, the supersymmetry breaking parameters are strongly
constrained by the necessity to avoid too large CP-violating and
flavor-violating effects in the $K$ and $B$ systems, in $\mu$ decays,
and in the electron and neutron EDMs. 
We therefore discuss in this paper whether a detectable
$d_\mu$ can arise in frameworks able to account for the constraints on
the structure of supersymmetry breaking, like a flavor blind mechanism
of SUSY breaking or some appropriate flavor symmetry. 
In the most
constrained situation with complex universal soft terms one has $A_\mu
\approx A_e$ and $m_{\tilde{e}}\approx m_{\tilde{\mu}}$ giving $d_\mu
\approx + d_e m_\mu/m_e$, necessarily a factor 3 below $10^{-24}\ecm$
given the present limit on $d_e$~\cite{deexp}.
More generically, it is
well known~\cite{dipoliSoft}
that a small phase in flavor-conserving soft terms
(like gaugino masses, $\mu$ and $B\mu$ terms)
can give a $d_e$ just below its experimental bound and a
$d_\mu\approx  d_e m_\mu/m_e$.
We will instead consider supersymmetric scenarios where
CP-violation must be accompanied by flavor violation
(similarly to what happens in the SM),
so that large CP-violating phases give acceptable dipoles that
evade the na\"{\i}ve scaling.

%

One way to account for the constraints on the flavor structure of
supersymmetry breaking is that the supersymmetry breaking mechanism
itself generates flavor-universal real soft terms at some scale $M_0$.
Even in this case, significant effects can arise if the theory above
some scale $\MT$ lower than $M_0$ contains additional sources of
CP-violation. Such effects leave their imprint in the soft terms
through radiative corrections arising between the scales $\MT$ and
$M_0$~\cite{HKR}.  Even if $\MT\gg M_Z$ and the source of CP-violation
decouples below $\MT$, CP-violation survives at lower
energies in the soft terms.  In section~\ref{GUT} we study the effects
due to SU(5) or SO(10) unification on a spectrum universal at the
Planck scale.  In section~\ref{nu} we consider the effects due to the
neutrino Yukawa couplings in the context of the see-saw mechanism.  In
section~\ref{both} we discuss unified see-saw models.
The muon EDM in
a left-right symmetric see-saw model has been studied in~\cite{BDM}.

In section \ref{flavor} we will consider the alternative possibility
that the same physics accounting for the structure of fermion masses
and mixings also determines the structure of sfermion masses.
Suitable flavor symmetries can force a non flavor-universal, but
phenomenologically acceptable, pattern of sfermion masses.  Even
assuming real $A$-terms (as suggested by the bounds on
EDMs~\cite{dNexp,deexp,dHgexp}), the complex lepton Yukawa matrix
gives rise to a non trivial pattern of CP-violating effects.

The results are summarized in section~\ref{conclusions}. A muon EDM
as large as $10^{-22\div 23}\ecm$ can be naturally obtained in a few
cases but none of the scenarios we consider guarantees that. 
On the other hand, an electron EDM
and a $\mu\to e \gamma$ rate within the sensitivity of planned
experiments~\cite{BH,DH,LFV} are a prediction of some of the scenarios
mentioned above.  If supersymmetry and $d_e$ were discovered, a measurement of
$d_\mu$ sufficiently accurate to distinguish $d_\mu/m_\mu \approx
+d_e/m_e$ from e.g.\ $d_\mu/m_\mu \approx -d_e/m_e$ would provide
interesting information on the source of the dipoles, and, eventually,
on the 11 element of the lepton mass matrix,
allowing to test interesting (but so far only theoretical) speculations about flavor.
Such a measurement needs a sensitivity to the muon EDM at least as low as $10^{-25}\ecm$.
In longer terms, this level of sensitivity could be reached using the
intense muon beam produced at a future
neutrino factory complex~\cite{dmu}.



\section{Effects from unified models}\label{GUT}

In unified models, flavor and CP violations cannot be confined to
quarks but must be present also in leptons, e.g. $b\to s \gamma$ must
be accompanied by $\mu\to e \gamma$.  In non supersymmetric GUT models
these leptonic effects are negligible because suppressed by powers of
$1/\MGUT$.  In supersymmetric GUT models with soft terms already
present above the unification scale, quantum corrections due to the
unified top quark Yukawa coupling imprint lepton flavor and CP
violations in the slepton mass terms~\cite{HKR}, inducing significant
computable effects in low energy processes~\cite{BH,DH}.

Minimal SU(5) unification induces an electric dipole of the up quark,
$d_u\propto\tan^4\beta$, of experimental interest if $\tan\beta$ is
large, but does not give detectable lepton electric
dipoles~\cite{vecchiDipoli}.  On the other hand, minimal SO(10)
unification gives $\mu\to e \gamma$ rates $(m_\tau/m_\mu)^2$-times
larger than in SU(5), and significant CP-violating
effects~\cite{DH,LFV}.  SO(10) effects in $\mu\to e \gamma$, $d_e$ and
$d_\mu$ are correlated by
\begin{equation}\label{eq:SO10dipoli}
d_e = \hbox{Im}\, d_{ee},\qquad
d_\mu = \hbox{Im}\, d_{\mu\mu},\qquad
{\rm BR}(\mu\to e\gamma) = \frac{m_\mu^3}{16\pi \Gamma_\mu}   (
|d_{e\mu}|^2 +|d_{\mu e}|^2) ,
\end{equation}
where
\begin{equation}\label{eq:dmatrix}
d_{\ell\ell'} = F \times \,V_{\ell_R \tilde{\tau}_R}V_{\ell'_L
\tilde{\tau}_L}V_{\tau_L \tilde{\tau}_L}^*V_{\tau_R 
\tilde{\tau}_R}^*
\end{equation}
so that $d_{e\mu}d_{\mu e} = d_{ee} d_{\mu\mu}$.  
The common loop factor $F$ is explicitly given in
appendix~\ref{GUTs}.
 In the same way as the CKM matrix measures
the flavor misalignement within the ${\rm SU}(2)_L$ multiplet of
left-handed up and down quarks, the $V_{\ell_L\tilde{\ell}_L}$
($V_{\ell_R\tilde{\ell}_R}$) mixing matrices measure the flavor
misalignement within the supermultiplet of left-handed (right-handed)
leptons and left-handed (right-handed) sleptons.  In the usual
supersymmetric flavor basis in which the charged lepton mass matrix is
diagonal, these matrices diagonalize the left-handed (right-handed)
slepton mass matrices.

\begin{table}[t]
$$\begin{array}{|rcllc|rcllc|} \hline
\multicolumn{4}{|c}{\hbox{\Blue present bound\Black}} & \hbox{\Blue future?\Black} &
\multicolumn{4}{c}{\hbox{\Blue present bound\Black}}  & \hbox{\Blue future?\Black}
\\ \hline 
d_N &<& 6.3~10^{-26}\ecm &\cite{dNexp}&     10^{-27}\ecm &
\hbox{BR}(\tau\to \mu \gamma) &<& 1.1~10^{-6}&\cite{expTauDec} & 10^{-9}
\\ 
d_e &<& 1.5~10^{-27}\ecm &\cite{deexp}&  10^{-29}\ecm &
\hbox{BR}(\mu\to e \gamma) &<& 1.2~10^{-11}&\cite{muegExp} &10^{-14}
\\ 
d_{^{199}\rm Hg} &<& 1.8~10^{-28}\ecm    &\cite{dHgexp}&  & 
\hbox{BR}(\mu\to e\bar{e}e) &<& 1.0~10^{-12}&\cite{mueExp} &10^{-16}
\\ 
d_\mu &<& 1.0~10^{-18}\ecm &\cite{dmuexp}&  10^{-25}\ecm &
\hbox{CR}(\mu\to e \hbox{ in Ti}) &<& 6.1~10^{-13}&\cite{expMuCapt} & 10^{-18}
\\
\hline
\end{array}$$\Black
\caption{\em Compilation of $90\%$ {\em CL} bounds on {\rm
CP}-violating and lepton-flavor violating processes.}
\label{tab:bounds}
\end{table}

Minimal SO(10) (defined as in~\cite{LFV}) predicts equal left and
right-handed lepton/slepton mixing, so that ${\rm BR}(\mu\to e\gamma)
= m^3_\mu|d_{ee}d_{\mu\mu}|/(8\pi\Gamma_\mu)$.  Furthermore, minimal
SU(5) and SO(10) models predict the unification of the
lepton and down-quark Yukawa matrices at $\MGUT$, which would imply
\begin{equation}\label{eq:VGUT}
|V_{e_L \tilde{\tau}_L}| = |V_{e_R \tilde{\tau}_R}| = |V_{td}|,\qquad 
|V_{\mu_L \tilde{\tau}_L}| = |V_{\mu_R \tilde{\tau}_R}| =| V_{ts}|\qquad
\hbox{renormalized at $\MGUT$}.
\end{equation}
However, it is well known that the corresponding minimal GUT relations
between $m_e,m_\mu,m_d,m_s$ are wrong by factors of $3$.  This problem
can be easily (and maybe even nicely~\cite{GJ}) solved in non minimal
GUTs\footnote{This implies that different entries of the Yukawa
matrices have different gauge structures, as given e.g.\ by Higgs
fields in bigger representations of the unified gauge group, or by
higher dimensional operators. These situations do not necessarily
imply non universal $A$-terms: computable non universal RGE
corrections~\cite{GG} to the $A$-terms get canceled by computable GUT
threshold effects~\cite{GUTdecoupling}, as dictated in a non trivial
way by supersymmetry.}
but in practice implies that relations\eq{VGUT} are also wrong by
unknown ${\cal O}(3)$ Clebsh factors.  Setting these factors to one we would
find
\begin{equation}\label{eq:relazioni0}
\frac{d_e}{d_\mu} \approx  \frac{V_{td}^2}{V_{ts}^2} 
\qquad\hbox{and}\qquad
d_\mu \circa{<} 10^{-25}\ecm
\sqrt{\frac{{\rm BR}(\mu\to e\gamma)}{10^{-11}}}.
\end{equation}
If this were the
case, minimal SO(10) effects would not generate a detectable $d_\mu$
given the bounds on $d_e$
(unless $d_{ee}$ has a small CP-violating phase) and on $\mu\to e
\gamma$.  The most recent bounds are collected in table~1.

Realistic relations between lepton-slepton mixing and CKM matrix can
only be obtained under assumptions on the lepton and down-quark Yukawa matrices.
A $d_\mu \circa{>}10^{-24}\ecm$ can arise
if the ${\cal O}(3)$ Clebsh factors suppress the 12 and 21 entries of the lepton Yukawa
matrix, so that $V_{e_L\tilde{\tau}_L}, V_{e_R\tilde{\tau}_R}$ are small.
However, 
the approximate unification of the down and lepton Yukawa matrices,
and the approximate equality $|V_{us}|\sim\sqrt{m_d/m_s}$
suggest a different assumption on the charged lepton Yukawa matrix~\cite{GJ}. 
It is interesting to consider the broad class of models 
with vanishing or sufficiently small
11, 13 and 31 entries of the charged lepton Yukawa matrix.
One gets
\begin{equation}\label{sys:relazioni}
\frac{d_e}{d_\mu} = \frac{d_{ee}}{d_{\mu\mu}} = -\frac{m_e}{m_\mu}\qquad\hbox{and}\qquad
d_\mu = 2.4\cdot 10^{-25}\ecm\;\sin\varphi_\mu 
\sqrt{\frac{{\rm BR}(\mu\to e\gamma)}{10^{-11}}}.
\end{equation}
where $\varphi_\mu$ is the phase of $d_{\mu\mu}$.  
Notice the opposite sign of $d_\mu/d_e$ with respect to the na\"{\i}ve
scaling relation, as discussed in greater detail in
section~\ref{flavor}.

\begin{figure}[t]\vspace{-8mm}
$$
\includegraphics[width=170mm,height=70mm]{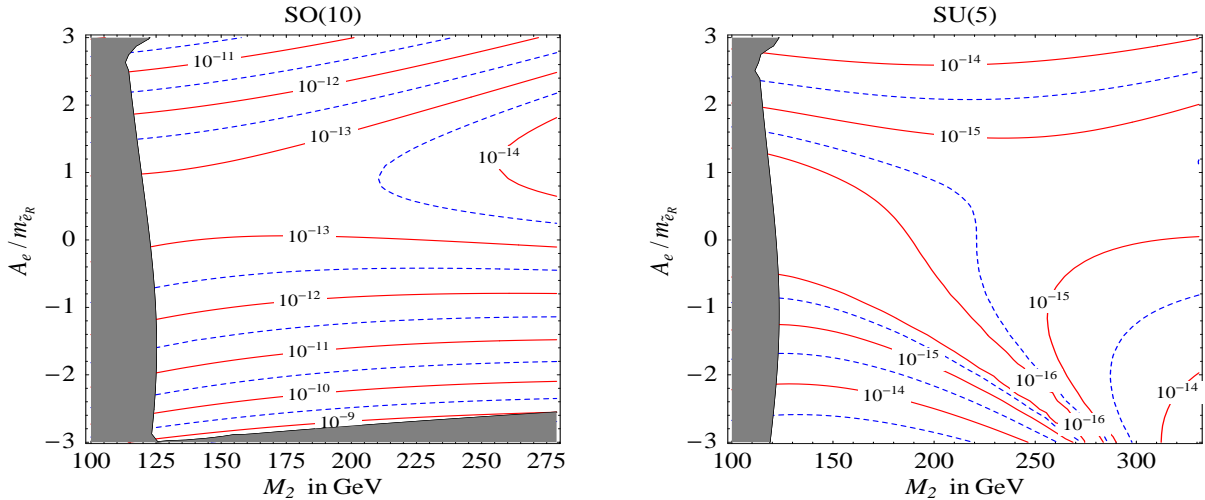}
$$
  \caption[]{\em Contour plot of $\hbox{\rm BR}(\mu\to e \gamma)$ in
${\rm SO(10)}$ (left) and ${\rm SU}(5)$ (right), for lepton-slepton
mixings as in eq.~(\ref{eq:VGUT}) and the parameter choice in
eq.\eq{parametri}.
\label{fig:mueg}}
\end{figure}

Unlike the ratio $d_e/d_\mu$, the values of $d_e$, $d_\mu$ and of the
$\mu\to e \gamma$ rate depend 
also on the structure of the 23 and 32 entries of the charged lepton and down quark Yukawa matrices.
In fig.~\ref{fig:mueg}a we show the prediction for ${\rm BR}(\mu\to
e\gamma)$ in the minimal SO(10) model assuming eq.\eq{VGUT}. 
The $\mu\to e\gamma$ rate obtained for any given assumption on the mixings,
and the corresponding predictions
for the muon and electron EDMs, can be then obtained 
by rescaling fig.~\ref{fig:mueg}a according to
eq.s~(\ref{eq:SO10dipoli}) and~(\ref{eq:dmatrix}). 
We also show the prediction for ${\rm BR}(\mu\to e\gamma)$ in the
minimal SU(5) model in fig.~\ref{fig:mueg}b, although, as remarked
above, SU(5) does not generate sizable $d_\mu$ and $d_e$.  Both
figures update the results obtained in~\cite{LFV}.  In the rest of
this section, we depart from the main theme of our work and discuss
the details of the computation.
The uninterested reader might want to jump to the next section.

\bigskip

An update of the computation in~\cite{LFV} is useful for various reasons.
The main reason is that the effects we are considering
strongly depend on the top Yukawa coupling at the GUT scale, see
eq.\eq{rescale}.  Before knowing the value of the top mass,
$\lambda_t(\MGUT)$ was estimated from theoretical prejudices about
proximity to an infrared-fixed point and about exact bottom/tau
unification~\cite{LFV}. There are other less relevant reasons to update
the computation. Charged sparticles lighter than $100\GeV$ have been
excluded by LEP.  In the MSSM, a small $\tan\beta\circa{<}2$ gives a
too light higgs.  In the SU(5) case, the computation in~\cite{LFV}
missed one diagram~\cite{missing2}: this gives an order one correction
(the correct expression is given in appendix~\ref{GUT}).  Finally,
assuming that the ``$g-2$ anomaly''~\cite{g-2} is due to supersymmetry
(rather than to QCD), it could be of some interest to study its
implications for other sleptonic penguin effects, like $\mu\to e
\gamma$ and the electric dipoles.

As in~\cite{LFV}, we assume minimal SU(5) and SO(10) unification
models, unified gaugino masses, universal scalar masses $m_0$ and
universal trilinear real $A$-terms $A_0$ at $M_{\rm Pl}
=2.4~10^{18}\GeV$.  The assumption of fully universal soft terms is
not demanded by experimental or theoretical requirements, but is often
employed in order to reduce the number of parameters (in our
computation it fixes the value of the $\mu$ term and of other less
relevant parameters).

RGE corrections in GUT models lead to non universal, flavor and
CP-violating soft terms at the Fermi scale, inducing observable
effects in leptons through one loop diagrams.  In SO(10) the effects
are dominated by electro-magnetic penguin diagrams.  They dominate
also in SU(5) if $\tan\beta\circa{>}{\rm few}$.  This implies
relations between different lepton flavor violating signals, like
\begin{equation}\label{eq:relations}
\frac{{\rm CR}(\mu\to e{\rm~in~Ti})}{{\rm BR}(\mu\to e\gamma)}\approx
0.5\cdot 10^{-2},\,\qquad 
\frac{{\rm BR}(\mu\to e\bar{e}e)}{{\rm BR}(\mu\to e\gamma)}\approx
0.7\cdot 10^{-2}.
\end{equation}
The correlation between $\mu\to e \gamma$ with $d_e$ and $d_\mu$
(previously discussed) and with $\tau \to \mu \gamma$ (see~\cite{LFV})
is different in the SU(5) and SO(10) cases.  The rates for all these
other processes can be
read from the $\mu\to e \gamma$ rate, that we plot in fig.\fig{mueg}a,b as
function of $M_2, A_e/m_{\tilde{e}_R}$.  We have assumed eq.\eq{VGUT}
and
\begin{equation}\label{eq:parametri}
\lambda_t(\MGUT) = 0.6,\quad\tan\beta=5,\quad m_{\tilde{e}_R}=300\GeV,\quad
|V_{ts}| = 0.04,\quad
|V_{td}|=0.01,\quad
\mu > 0
\end{equation}
Unless otherwise indicated all parameters are renormalized at the
electroweak scale.  For any other value of the parameters ${\rm
BR}(\mu\to e \gamma)$ can still be read from fig.\fig{mueg} because it
approximatively scales as
\begin{equation}\label{eq:rescale}
\hbox{BR}(\mu\to e \gamma)\propto 
\lambda_t^p(\MGUT) \times m_{\tilde{e}_R}^{-4} \times (\mu
\tan\beta)^2\times\left\{\begin{array}{ll} 
|V_{e_R \tilde{\tau}_R} V_{\mu_R \tilde{\tau}_R}|^2 &\hbox{in SU(5)}\cr
|V_{e_R \tilde{\tau}_R} V_{\mu_L \tilde{\tau}_L}|^2+|V_{e_L
\tilde{\tau}_L} V_{\mu_R \tilde{\tau}_R}|^2 &\hbox{in SO(10)}\cr 
\end{array}\right.
\end{equation}
where $p=4$ in SU(5) and $p=8$ in SO(10).  When rescaling
$m_{\tilde{e}_R}$ one has also to rescale the other mass parameters,
$M_2$ and $A_e$.  This na\"{\i}ve rescaling is a good approximation,
unless sparticle masses are comparable to the $Z$ mass.  The upper
bound on $M_2$ in fig.\fig{mueg}a,b corresponds to the assumption
$m_0^2>0$.  A large $\tan\beta$ can be naturally obtained from a small
$\mu\propto 1/\tan\beta$, or by accidental cancellations.  We have
assumed a positive $\mu$ term, as suggested by data about $b\to s
\gamma$ (and $g-2$, if attributed to supersymmetric effects), and
computed it assuming an universal scalar mass term $m_0$.  For this
choice of $\mu$ there can be accidental cancellations between the
Feynman diagrams that contribute to the $\mu\to e \gamma$ rate in
SU(5) (see fig.\fig{mueg}b).

\begin{floatingfigure}[r]{8cm}
$$
\includegraphics[width=70mm,height=60mm]{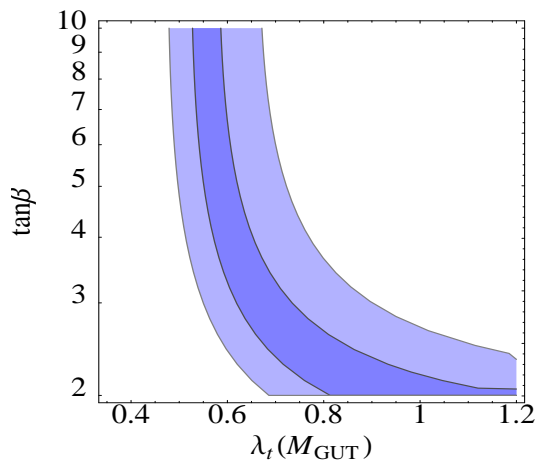}$$
  \caption[]{\em Values of $\lambda_t(\MGUT)$ compatible with  $M_t=(175\pm 5)\GeV$ (outer band) and
with $M_t = 175\GeV$ (inner band).
\label{fig:top_GUT}}
\end{floatingfigure}
In order to do a precise computation we fixed the value of
$\lambda_t(\MGUT)$, rather than the value of $M_t$.  Fig.\fig{top_GUT}
(from~\cite{largem0}) shows the value of $\lambda_t(\MGUT)$ extracted
from the measured pole top mass
$$M_t=(175\pm 5) \GeV.$$ The values of $\lambda_t(\MGUT)$ are somewhat
smaller than the ones used in~\cite{LFV}, if values of
$\tan\beta\circa{<}2$ are excluded because give a too light Higgs mass
(as happens in the MSSM, unless one adds extra singlets).  Decreasing
$\lambda_t(\MGUT)$ by a factor 2 reduces the SU(5) (SO(10)) prediction
for the $\mu\to e \gamma$ decay rate by one (two) orders of magnitude.
The dependence of the predicted $\mu\to e \gamma$ rate on $\tan\beta$ is not only due to the
explicit $\tan^2\beta$ factor in eq.\eq{rescale}, but also to the dependence 
of the experimental band for $\lambda_t(\MGUT)$
on $\tan\beta$ depicted in fig.\fig{top_GUT}.
As a consequence, the $\mu\to e\gamma$ rate
is minimal around
$\tan\beta\sim 4$.

The width of the range for $\lambda_t(\MGUT)$ is not only due to the
few $\%$ uncertainty on $M_t$, but also to unknown sparticle threshold
corrections affecting the value of $\lambda_t$ just above the SUSY
breaking scale. Moreover, the running up to $\MGUT$ depends on the
gauge couplings (also affected by unknown sparticle threshold
corrections) and amplifies the uncertainties in $\lambda_t$.
In fact, the RGE evolution of $\lambda_t$ exhibits an infra-red fixed
point behavior: different values of $\lambda_t(\MGUT)$ converge in a
restricted range of values of $\lambda_t$ at low energy~\cite{IR}.
This allowed to guess the value of $M_t$ from assumptions about
$\lambda_t(\MGUT)$~\cite{IR}.  Now that $M_t$ is known we would like
to do the opposite and renormalize $\lambda_t$ from low to high
energies.  We then see the reverse of the medal: $\lambda_t(\MGUT)$ is
not strongly constrained by the measured value of $M_t$.  Even
assuming an exactly known $M_t$, sparticle threshold effects would
still induce a large uncertainty in $\lambda_t(\MGUT)$ (as shown by
the inner band in fig.\fig{top_GUT}).  It is useful to remember that
in the pure SM the theoretical error on the relation between $M_t$ and
$\lambda_t(M_t)$ (NNLO QCD corrections have not been computed) is
equivalent to a $\pm 2\GeV$ uncertainty in $M_t$.

\medskip

One can wonder if attributing the $g-2$ anomaly to supersymmetry
(rather than to QCD effects) would allow to constrain the predictions
for the EDMs and $\mu\rightarrow e\gamma$ by reducing the uncertainty
on the values of $\mu,\tan\beta$ and of the slepton masses.  In fact
the one loop diagrams that contribute to $\mu\to e \gamma$ and to
$g-2$ are quite similar.  This analysis was done in~\cite{Hisano} for
the case discussed in the next section, but is particularly
interesting in the case of the GUT-induced effects, where the
uncertainty on sparticle masses induces the dominant uncertainty on
the $\mu\to e \gamma$ rate.  We find a strong correlation only if we
assume that $m_0=A_0=0$ at $M_{\rm Pl}$ \footnote{This `no
scale'~\cite{noscale} boundary condition is probably the simplest way
of justifying universal sfermion masses.}, in which case the $A$-terms
and sfermion masses are induced by radiative corrections proportional
to the gaugino masses. 
In this case we find
\begin{equation}\label{eq:correlazione}
{\rm BR}(\mu\to e \gamma) \approx\left\{
\begin{array}{ll}
\displaystyle
3~10^{-13} \lambda_t^4(\MGUT) \bigg(\frac{\delta a_\mu}{4~10^{-9}}\bigg)^{\!2}
\left|\frac{V_{e_R \tilde{\tau}_R} V_{\mu_R \tilde{\tau}_R}}{0.01\cdot
0.04}\right|^2  &\hbox{ in SU(5)} \\[3mm] 
\displaystyle
2~10^{-11} \lambda_t^8(\MGUT) \bigg(\frac{\delta a_\mu}{4~10^{-9}}\bigg)^{\!2}
\frac{|V_{e_R \tilde{\tau}_R} V_{\mu_L \tilde{\tau}_L}|^2+
|V_{e_L \tilde{\tau}_L} V_{\mu_R \tilde{\tau}_R}|^2}{2(0.01\cdot
0.04)^2} & \hbox{ in SO(10)} 
\end{array}
\right.,
\end{equation}
where $\delta a_\mu$ represents the supersymmetric contribution to
$a_\mu = (g-2)/2$.  In the general case, there is only a loose
correlation between $\mu\to e \gamma$ and $\delta a_\mu$.  One reason
is that the GUT-induced $\mu\to e \gamma$ rate depends on how much RGE
effects due to the unified top Yukawa coupling make the staus lighter
than selectrons and smuons: the amount of non-degeneracy depends on
$m_0$ and $A_0$ and is minimal at $m_0 = A_0 = 0$.  Therefore, in the
SO(10) case, eq.\eq{correlazione} provides a lower bound on the
$\mu\to e \gamma$ rate.  In the SU(5) case there can be accidental
cancellations between the Feynman diagrams that contribute to $\mu\to
e \gamma$ (see fig.\fig{mueg}b): if $m_0,A_0\neq0$ the $\mu\to e
\gamma$ rate can be above or below the value in\eq{correlazione}.

\section{Effects from neutrinos}\label{nu}
Similar effects can be generated by the neutrino Yukawa couplings present in
supersymmetric see-saw models.  Adding to the MSSM ``right-handed
neutrinos'' $N_i$, the most generic lepton superpotential
\begin{equation}\label{eq:seesawmass}
\mathscr{W} = \frac{M_{ij}}{2}N_iN_j+\lambda_N^{ij}\, L^iN^j H_{\rm u} +
\lambda_E^{ij} \,E^i L^j  H_{\rm d},
\end{equation}
gives the Majorana neutrino masses
\begin{equation}\label{eq:seesaw}
 \mb{m}_\nu=  -\mb{\lambda}_N\cdot \frac{v^2
 \sin^2\beta}{\mb{M}}\cdot \mb{\lambda}_N^T 
\end{equation}
(when appropriate we use boldface to emphasize the matrix structure).
Trough the same mechanism operative in GUT models, RGE effects
proportional to the squared Yukawa coupling of the right-handed
neutrinos imprint CP and lepton flavor violations in slepton
masses.
For example, the correction to the $3\times 3$ mass matrix of
left-handed sleptons is
\begin{equation}\label{eq:slepton}
\delta \mb{m}^2_{\tilde{L}} = -\frac{1}{(4\pi)^2} (3m_0^2 + A_0^2)
\mb{Y}_N +\cdots,\qquad\hbox{where}\qquad 
\mb{Y}_N\equiv \mb{\lambda}_N^*
\ln(\frac{\MGUT^2}{\mb{MM}^\dagger})
\mb{\lambda}_N^{T}
\end{equation}
having assumed universal soft terms at $\MGUT$ and neglected ${\cal
O}(\lambda_N^4)$ effects.  In this approximation, the experimental
bounds from $\ell_i\to \ell_j \gamma$ decays are saturated for 
\begin{equation}\label{eq:Ynu}
[\mb{Y}_N]_{\tau \mu} ,[\mb{Y}_N]_{\tau e} \sim 10^{1\pm 1},\qquad
[\mb{Y}_N]_{\mu 
e} \sim 10^{-1\pm 1}\end{equation}
having assumed values of sparticle masses compatible with a supersymmetric
explanation of the $g-2$ anomaly.
A more complete
analysis can be found in~\cite{LMS}.

Since $\mb{M}$ is unknown, the see-saw relation\eq{seesawmass} does
not allow to convert the measurement of the neutrino masses
$\mb{m}_\nu$ into useful restrictions on the scale of
$\mb{\lambda}_N$, or on its flavor structure. In particular, the
large $\nu_\mu/\nu_\tau$ mixing observed in atmospheric
oscillations~\cite{SKatm} does not necessarily imply a correspondingly
large SUSY mixing in slepton interactions.  The
$\mb{M},\mb{\lambda}_N$ and $\mb{\lambda}_E$ matrices that describe
the supersymmetric see-saw superpotential of eq.\eq{seesawmass}
contain 15 real parameters and 6 CP-violating phases.  At low energy, in
the mass eigenstate basis of the leptons, 3 real parameters describe
the lepton masses, and both the neutrino and the left-handed slepton
mass matrices are described by 6 real parameters and 3 CP-violating phase~\cite{DI}.  
Since $(15+6) = (3+0) + (6+3)+(6+3)$ we see 
that see-saw mechanism has too many free parameters to allow to make general predictions:
{\em any
pattern of lepton and neutrino masses is compatible with any pattern of
radiatively-generated flavor violations in left-handed slepton
masses}\footnote{Since there are no restrictions, it is of course
possible to obtain values of $\hbox{BR}(\mu\to e \gamma)$ or of
$\hbox{BR}(\tau\to \mu \gamma)$ just below their experimental bounds.
This is necessarily the case, under appropriate assumption on the size
of the neutrino Yukawa couplings~\cite{Hisano2,Hisano} eventually
justified by choosing some flavor model~\cite{SubitoSotto}. 
On the contrary, too large $\mu\to e \gamma$ rates are normally obtained under other
reasonable assumptions not demanded by experimental data, as discussed in~\cite{CI}.
In general,
eq.s\eq{Ynu} tell that a significant improvement of the experimental
sensitivity in $\tau\to \mu\gamma$ would allow to test if the neutrino
Yukawa matrix contains order one entries.}. 
The RGE effects in $A$-terms can be predicted in terms of the RGE effects in left-handed slepton masses.
Unlike in the GUT case, it is not possible to predict the relative
size of the different flavor and CP violating processes like $\mu\to e
\gamma$, $\tau\to \mu\gamma$, $d_e$, $d_\mu$ and to assess which one
has a better chance of being observed, if any. For these reasons we
will not perform `detailed' computations.

\medskip

In view of this situation, one can try to see if useful informations
can be obtained from models of fermion masses, rather than directly
from fermion masses.  In this respect, it is important to appreciate
that the requirement of having a large atmospheric mixing angle
between the {\em most splitted\/} neutrino states ($\Delta m^2_{\rm
atm}\gg \Delta m^2_{\rm sun}$) gives significant
restrictions\footnote{Experimental data prefer, but do not require,
$\Delta m^2_{\rm sun}\ll\Delta m^2_{\rm atm}$. 
If this were not the case,
neutrino data would not give significant restrictions on flavor
models.  Furthermore we restrict our attention on models that predict
$\theta_{\rm atm}\sim 1$. Depending on personal taste, one could
instead be content with models that predict $\theta_{\rm atm}\sim
\epsilon^{1/2}$ (where $\epsilon$ is a relatively small number), or
want models that predict $\theta_{\rm atm}$ close to $\pi/4$.},
suggesting two peculiar structures for the Majorana neutrino mass matrix
$\mb{m}_\nu$: in the limit 
$\Delta m^2_{\rm sun} = 0$ they are
\begin{itemize}
\item[(a)]  a rank one matrix (if $\Delta m^2_{\rm atm}>0$, i.e.\ if neutrinos have a hierarchical spectrum);
\item[(b)]  a rank two pseudo-Dirac matrix (if $\Delta m^2_{\rm atm}<0$, i.e.\
if neutrinos have an inverted spectrum)\footnote{Neutrinos 
could also have a degenerate or quasi-degenerate spectrum, but we are not interested in
these cases.
We also do not consider inverted spectra
with a generic Majorana phase incompatible with a
pseudo-Dirac structure, that automatically guarantees
the smallness of $\Delta m^2_{\rm sun}/\Delta m^2_{\rm atm}$ and makes it
stable under radiative corrections.}.
\end{itemize}
Assuming $\theta_{\rm atm}=\pi/4$ and
$\theta_{\rm CHOOZ} = 0$ (i.e.\ maximal
$\nu_\mu \leftrightarrow \nu_\tau$ atmospheric oscillations), 
in the mass eigenstate basis of charged leptons, these matrices can be explicitly written as
$$\mb{m}_\nu({\rm a}) \propto  
\begin{pmatrix}
0&0&0 \cr 0&1&1\cr 0&1&1
\end{pmatrix},\qquad 
\mb{m}_\nu({\rm b})\propto
\begin{pmatrix}0&1&1 \cr 1&0&0\cr 1&0&0
\end{pmatrix}.$$
In the see-saw context, these mass
matrices are generated by the following
superpotentials~\cite{King,nuTextures}
\begin{equation}\label{eq:seesawmotivato}
\mathscr{W}({\rm a})= \lambda N (L_\mu + L_\tau)H_{\text{u}} + \frac{M}{2} N^2,\qquad
\mathscr{W}({\rm b}) =\lambda N (L_\mu + L_\tau)H_{\text{u}} +
\lambda' N' L_eH_{\text{u}} + M  NN'
\end{equation}
that could be both justified by a broken
$L_e-L_\mu-L_\tau$ symmetry (that would suppress $\mu\to e \gamma$).
Having reproduced the main structure of the neutrino mass matrix, it
is easy to add the solar mass splitting (large solar mixing is
automatically obtained in case b), and to inglobate neutrino masses in
a full model of fermion masses.  In conclusion, the only generic
suggestion from neutrino data is that one right handed
neutrino mass eigenstate has comparable Yukawa couplings to $\tau$ and
$\mu$ and a smaller coupling to $e$.  What are the implications
of this restricted structure for supersymmetric lepton-flavor and
CP-violating effects?  The Yukawa coupling $\lambda$ is the crucial
parameter for supersymmetric effects, but its size cannot be deduced from the see-saw relation in eq.\eq{seesaw}.  
Concrete examples that
illustrate how both large and small SUSY mixings are compatible with
this restricted see-saw structure have been presented in~\cite{LMS}
(for case (a) --- the extension to case (b) is trivial).  In fact, the
see-saw structure\eq{seesawmotivato}
does not need large mixing angles
in any Yukawa matrix (if $\lambda$ is not the largest element of $\mb{\lambda}_N$:
see~\cite{AFM} for explicit examples). 
The casistics discussed
in~\cite{LMS} contains all what can be reliably said about
neutrino-Yukawa-induced $\mu\to e \gamma$ and $\tau\to \mu \gamma$
decays.

\medskip

In the assumption that
$\lambda$ is large enough (e.g.\ $\lambda\sim \lambda_t$),
the minimal see-saw structure of eq.\eq{seesawmotivato} generates
significant effects in the $\tau,\mu$ sector (e.g.\ the $\tau\to \mu
\gamma$ decay, see eq.\eq{Ynu}) but does not generate a muon
electric dipole.
In fact, by redefining the phases of left and
right-handed leptons, CP violation can be rotated away from this large
Yukawa coupling $\lambda$ and confined to the smaller Yukawa couplings
of the other right-handed neutrinos.

\medskip

Is it possible
to obtain sizable lepton EDMs by choosing appropriate values for the many unknown see-saw parameters
(compatible with neutrino masses, but not suggested by them)?
We need to further assume that many entries of $\mb{\lambda}_N$ are
large, with the hierarchy in neutrino masses obtained from a hierarchy
in the masses $\mb{M}$.  The computation of electric dipole moments is
tricky, but general arguments simplify in a useful way the
analysis. 
(We remind that here we assume that soft terms are universal at $\MGUT$:
the more interesting case of unified see-saw models with universal soft terms at $M_{\rm Pl}$
will be considered in the next section). 
Then the $3\times 3$ flavor matrix of charged lepton
electro-magnetic moments $\mb{d}$ (see
eq.s~(\ref{eq:SO10dipoli}) or~(\ref{eq:AmuSU5})) is restricted by the ${\rm
U}(3)_L\otimes{\rm U}(3)_E\otimes{\rm U}(3)_N$ symmetry of the flavor-universal 
part of the Lagrangian and by the holomorphicity of
supersymmetry.  The contribution to $\mb{d}$ that dominates the EDMs
has the form\footnote{For brevity, here we only give the final result.
The use of U(3) symmetries is explained in~\cite{vecchiDipoli}.  The
consequences of the holomorphicity of supersymmetry for soft breaking
terms have been nicely described in~\cite{GR}. Alternatively,
the way in which the factors ${\lambda}_N$ and ${M}$ appear can be
understood by inspecting the explicit form of the RGE equations.}
\begin{equation}\label{eq:dominant}
\mb{d} \propto  \mb{\lambda}_E\cdot \mb{Y}_N\cdot \mb{Y}_E\cdot
\mb{Y}_N^2\qquad\hbox{where}\qquad
\mb{Y}_E \equiv \mb{\lambda}_E^\dagger \cdot \mb{\lambda}_E
\end{equation}
and $\mb{Y}_N $ is defined in eq.\eq{slepton}.
The muon electric dipole is the imaginary part of the $\mu\mu$
diagonal element of the matrix $\mb{d}$, as in eq.\eq{SO10dipoli}.
Eq.\eq{dominant} gives electric dipoles $d_e$ and $d_\mu$ suppressed
by $\lambda_\tau^2$, so that large dipoles can be obtained at large
$\tan\beta$:
\begin{equation}
\label{seesawSM}
d_\ell \sim  \frac{e\alpha_{\text{em}}}{4\pi}\frac{m_\ell}{\tilde
m^2_{\tilde{\ell}}} \lambda_{N_3}^4 \lambda_{N_2}^2 \lambda_\tau^2 J'_{\text{CP}} 
\end{equation}
where  $\lambda_{N_{3}}$, $\lambda_{N_{2}}$ are the largest
and next-to-largest neutrino Yukawa eigenvalues and $J'_{\text{CP}}$
is the Jarskog invariant associated to the matrices $\mb{Y}_N$ and $\mb{\lambda}_E$.
If $\lambda_{\nu_{3}}\sim \lambda_{\nu_{2}}\sim 1$ large lepton EDMs can be generated.
However, whatever is the flavor structure of $\mb{\lambda}_N$, the electric
dipoles satisfy the relation $d_\mu/m_\mu\approx - d_e/m_e$ (as can be
understood by noticing that $d_\tau/m_\tau$ and $\sum_\ell
d_\ell/m_\ell\propto \hbox{Im Tr}(\mb{Y}_N\cdot \mb{Y}_E\cdot
\mb{Y}_N^2) $ are small~\cite{vecchiDipoli}). 
The na\"{\i}ve scaling between $d_e$ and $d_\mu$ can be evaded using the flavor structures
that only appear at subleading orders (e.g. two loop RGE), different
from $\mb{Y}_N$ because contain a smaller power of
$\ln(\MGUT^2/\mb{MM}^\dagger)$.
Since large neutrino couplings $\lambda_N^{ij}$ imply $M_{ij}$ not much smaller than $\MGUT$,
the $d_\mu$ generated by `subleading' radiative effects could be sizable and larger than the one
generated by `leading' radiative effects.  The most optimistic thing that can be said about this
possibility is that it is not excluded.

\section{Effects from neutrinos and from SU(5)}\label{both}

A more appealing possibility arises if the gauge structure at the
scale at which the soft terms are universal is richer than the SM
one. As seen in section 2, an enhancement of LFV processes and EDMs
requires non-universality in both the right-handed and left-handed
slepton sectors. RGE corrections due to right handed neutrinos do not
affect right handed sleptons, and RGE effects do to the unified Yukawa
of the top in SU(5) GUTs do not affect left-handed sleptons.  If both
neutrino and SU(5) effects are present, the amplitudes of penguin
diagrams (that give EDMs and $\ell_i\to \ell_j \gamma$ decays) get
enhanced by $m_\tau/m_\mu$ factors as in SO(10).  The relevant Yukawa
interactions are described by the superpotential
$$ \mathscr{W}=\lambda_U^{ij} T_iT_j H + \lambda_{DE}^{ij}
F_iT_j\bar{H} + \lambda_N^{ij} F_i  N_j H  $$ 
where $T$ ($F$) are the usual SU(5) 10-plets
($\bar{5}$-plets) and $H,\bar{H}$ are the SU(5) Higgs fields.  It is
convenient to assume that $\mathscr{W}$ is written in the
mass-eigenstate basis of the right-handed neutrino singlets $N_j$.  As
discussed in the previous section, the atmospheric neutrino anomaly
and the CHOOZ bound motivate the assumption that one of the
right-handed neutrinos (say, $N_3$) has comparable couplings to the
second and third generation and a smaller coupling to the first
generation:
$$\lambda_N^{i3}\equiv U_{i3} \lambda,\qquad\hbox{with}\qquad
U_{\tau 3}\sim U_{\mu 3}\gg U_{e3}\qquad\hbox{and}\qquad U_{e3}^2
+U_{\mu 3}^2 +U_{\tau 3}^2 =1.$$ 
We now show that a large $d_\mu$ can be obtained in this minimal
unified see-saw context, provided that $\lambda\approx \sqrt{M/10^{15}\GeV}$ is large enough.  As
previously discussed, this requires to
assume an appropriate value, around the GUT scale, for the mass $M$ of the right-handed neutrino $N_3$.

Since we are interested in the electric dipoles of the charged leptons
we can neglect the small up and charm couplings in the up-quark
matrix, $\lambda_c = \lambda_u = 0$.  In this unified see-saw context we can also
conservatively assume that the other neutrino Yukawa couplings are
small and neglect them: $\lambda_N^{i2}=\lambda_N^{i1}=0$.  The
important point is that in this relevant limit it is no longer
possible to rotate away CP-violation from the relevant Yukawa
interactions
$$ \mathscr{W}=\lambda_{t} T_3 T_3 H + \lambda_{DE}^{i3} F_i
T_3\bar{H} + \lambda_N^{i3} F_i  N_3 H. $$ 
The phases in $\lambda_t$, in $\lambda_{DE}^{i3}$
and one of the phases in $\lambda^{i3}_N$ can be rotated
away\footnote{In order to rotate away one of the phases in
$\lambda^{i3}_N$ one has to redefine the $N_3$ field,
eventually giving a complex right-handed neutrino mass term $M$. 
However, the effects we are considering only depend on $M^\dagger M$, so that a complex $M$ 
is indeed irrelevant.} by redefining
the phases in
$T_3, F_i,N_3$,
but the remaining phases in $\lambda_N^{i3}$ are physical.  They
induce the following EDMs
$$ d_e   \sim \frac{\alpha_{\rm em}  m_\tau }{4\pi
m_{\tilde{\tau}}^2}\lambda^2 \lambda_t^2  \hbox{Im}\,V_{td}  U_{e 3}
(U_{\tau 3} 
V_{tb} )^*,
\qquad
d_\mu \sim \frac{\alpha_{\rm em}  m_\tau }{4\pi
m_{\tilde{\tau}}^2}\lambda^2 \lambda_t^2 \hbox{Im}\,  V_{ts}  U_{\mu
3} (U_{\tau 3} 
V_{tb} )^*,$$ 
where we have assumed
eq.s~(\ref{eq:VGUT}). Therefore, $d_\mu$ can be large and much larger
than $d_e$.  Using the techniques described in the previous section,
the same result can be reobtained in a more elegant way by noticing
that the contribution to $\mb{d}$ relevant for the EDMs is
proportional to
$\mb{\lambda}_U^{\phantom{\dagger}}\mb{\lambda}_U^\dagger
\mb{\lambda}_{DE}^{T\phantom{\dagger}}
\mb{\lambda}_N^{*\phantom{\dagger}}
\mb{\lambda}_N^{T\phantom{\dagger}}$.  
We
now discuss the bounds on $d_\mu$ set by other processes.  The
strongest bound comes from $\tau\to \mu\gamma$ and depends on the
details of the model.  Omitting unknown order one factors and assuming
large CP-violating phases we estimate\footnote{Bounds from $\tau\to
\mu \gamma$ have been discussed in~\cite{FMS} in a more general
context but using the `mass-insertion approximation'.
An analogous analysis has been performed in~\cite{BS} in the quark sector.}
$$d_\mu \sim \hbox{few}\cdot 10^{-23} \,e\,{\rm
cm}\frac{|V_{ts}|}{0.04}\sqrt{\frac{\hbox{BR}(\tau\to\mu
\gamma)}{10^{-6}}}.$$ 
Since $d_\mu \circa{<} d_eV_{ts}/V_{td}U_{e3}$
the experimental bound on $d_e$ 
induce an upper bounds on $d_\mu$.
The upper bound on $d_\mu$ induced by the experimental bound on $\mu\to e \gamma$ is comparable and
also dependent 
on the unknown parameter $U_{e3}$.   As remarked above, the parameter $U_{e3}$ is
not directly related to the corresponding measurable element of the
neutrino mixing matrix: $U_{e3}$ can be naturally smaller than
 $\theta_{\rm CHOOZ}$
(that can be generated by another right-handed neutrino).  The
opposite possibility, $\theta_{\rm CHOOZ}\ll U_{e3}$, would instead
need accidental cancellations.  Within the minimal see-saw structure
suggested by neutrino data, eq.\eq{seesawmotivato}, the additional
assumption that the lepton Yukawa matrix has a vanishing 11 element
and comparable 12 and 21 entries gives rise to a relatively large
$U_{e3} \sim \sqrt{m_e/2m_\mu}$ (that will be
experimentally tested in long-baseline neutrino experiments) and
consequently to $d_\mu/m_\mu\sim d_e/m_e$.

\section{Flavor symmetries}\label{flavor}
We now consider the possibility that the structure of soft terms is
constrained by the same physics giving rise to the pattern of fermion
masses and mixings.  A generic flavor symmetry that explains the
observed pattern of fermion masses does not necessarily force an
acceptable pattern of sfermion masses. For example, U(1) symmetries do
not relate the diagonal soft mass terms, all separately invariant
under the symmetry. On the other hand, the required approximate
degeneracy of the first and second family sfermion masses can be
guaranteed by a suitable non-abelian symmetry. It is then interesting
that the simplest non-abelian symmetry accounting for the smallness of
the two lighter fermion families masses also automatically guarantees
this approximate degeneracy~\cite{u2}.  
In this section, we discuss the expectations for the muon and electron
EDMs in this context. 

Rather than focusing on a particular model, we would like to study the
generic properties of this class of models.  In order to simplify the
discussion we will make few assumptions, valid in wide classes of
models.  Generically, the lepton Yukawa matrix and the slepton masses
can be diagonalized by unitary matrices $U_L,U_R,T_L,T_R$ as
\begin{equation}
\label{eq:rotazioni}
\lambda_E = U^\dagger_R \lambda_{E,\text{diag}} U_L, \qquad
m_L^2 = T_L^\dagger m^2_{L,\text{diag}} T_L ,\qquad
m_{e_R}^2 = T_R^\dagger m^2_{e_R,\text{diag}} T_R \; .
\end{equation}
For a naturally hierarchical Yukawa matrix $\lambda_E$, the mixing
matrices $U_{L,R}$ can be written with sufficient accuracy as the
product of three $2\times 2$ rotations diagonalizing sector by sector
the Yukawa matrix in subsequent steps: $U_L = U^L_{13} U^L_{12}
U^L_{23}$, $U_R = \Phi\, U^R_{31} U^R_{21} U^R_{32}$, where
$\Phi=\diag(e^{i\phi_e},e^{i\phi_\mu},e^{i\phi_\tau})$ removes the
phases from the diagonal elements one is left with after the three
rotations. Since the Yukawa matrices are in general complex, the
``$ij$'' rotations also involve a phase in their off diagonal
elements.


The first assumption we make is that the 13 rotations are negligible.
This is the case if the 13 and 31 entries of the charged lepton Yukawa
matrix are sufficiently small. In the quark sector, the same
assumption on the quark Yukawa matrices is supported by the pattern of
quark masses and mixings. In fact, the presence of approximate
``texture zeros'' in the 13, 31 positions, together with the presence
of an approximate zero in the 11 position and the approximate equality
in magnitude of the 12 and 21 elements, successfully accounts for the
``leading order''~\cite{RRRV} relations between quark masses and
angles $|V_{ub}/V_{cb}|\sim \sqrt{m_u/m_c}$, $|V_{td}/V_{ts}|\sim
\sqrt{m_d/m_s}$~\cite{HR}.  Here, however, we will only make the
assumption that the 13 and 31 elements of $\lambda_E$ are negligible.
Since the Yukawa matrix and the corresponding $A$-term matrix have the
same quantum numbers under the flavor symmetry determining their
structure, the $A$-terms can be written as $\A_{ij} =
A_{ij}\lambda_{ij}$, where the various $A_{ij}$ are naturally
comparable but not equal.

Secondly, we assume that effects due to non degeneracy between
selectrons and smuons can be neglected.  Any successfully broken
flavor symmetry must guarantee that degeneracy at a high level of
accuracy.  Finally, we again assume that 13 rotations can be neglected
also in the slepton sectors.  In conclusion, the left and right-handed
slepton mass matrices can be diagonalized by a rotation in the ``23
sector'' only.  If the rotation in the 23 sector is not too large,
this hypothesis takes under control potentially dangerous FCNC and
CP-violation effects.

To summarize, we assume the following structures
\begin{equation}\label{eq:iniziale}
\lambda = 
\begin{pmatrix}
\lambda_{11}& \lambda_{12}&0 \cr
\lambda_{21}  &  \lambda_{22}  & \lambda_{23}  \cr
0 & \lambda_{32}  &\lambda_{33}
\end{pmatrix},\qquad
\tilde{A}=A\lambda = \begin{pmatrix}
A_{11}\lambda_{11}& A_{12}\lambda_{12}&0 \cr
A_{21}\lambda_{21}  & A_{22} \lambda_{22}  & A_{23}\lambda_{23}\cr
0 & A_{32}\lambda_{32} &A_{33}\lambda_{33}
\end{pmatrix},\qquad
m^2 = \begin{pmatrix}
m^2_{1,2}& 0&0 \cr
0  & m^2_{1,2}  &m^2_{23}  \cr
0&m^2_{23}   &m^2_3
\end{pmatrix}
\end{equation}
for the various Yukawa, $A$-term\footnote{A similar structure of
$A$-terms has been considered in~\cite{MM}.} and sfermion mass
matrices.

\begin{figure}[t]
\setlength{\unitlength}{1mm}$$
\begin{picture}(160,25)
\put(0,3){\includegraphics[width=157mm]{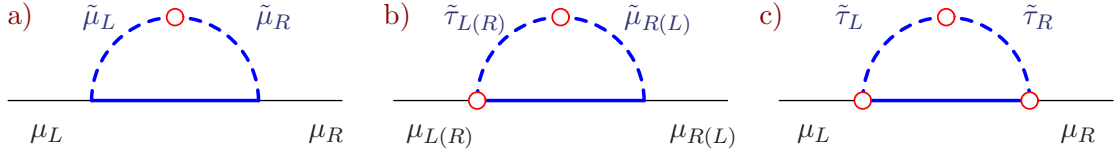}}
\put(8,0){$\mu_L$}\put(45,0){$\mu_R$}\Blue\put(15,15){$\tilde{\mu}_L$}\put(38,15){$\tilde{\mu}_R$}\Black
\put(58,0){$\mu_{L(R)}$}\put(93,0){$\mu_{R(L)}$}\Blue\put(63,15){$\tilde{\tau}_{L(R)}$}\put(87,15){$\tilde{\mu}_{R(L)}$}\Black
\put(110,0){$\mu_L$}\put(145,0){$\mu_R$}\Blue\put(115,15){$\tilde{\tau}_L$}\put(140,15){$\tilde{\tau}_R$}\Black
\Red\put(5,15){a)}\put(55,15){b)}\put(105,15){c)}\Black
\end{picture}$$
  \caption[]{\em The three type of contributions to $d_\mu$. The photon (not shown) should be coupled
to charged particles.
\label{fig:dipoli}}
\end{figure}

We are now ready to compute the EDMs\footnote{The supersymmetric
flavor problem can be partially alleviated by alternatively assuming
that the sfermions of the first two generations have few TeV mass.
This possibility would give similar EDMs as the one we consider
here.}.  Let us first consider the muon EDM. 
As discussed in appendix~A, eq.s~(\ref{dmufirstorder},\ref{dmucontr}),
the general one-loop expression for
$d_\mu$, eq.\eq{dmugeneral}, can be conveniently simplified as a sum of three dominant
contributions, pictorially represented in fig.\fig{dipoli}.
The first diagram is proportional to $\tilde{A}_{\mu_L \mu_R}$, the 22
element of the $A$-term matrix, written in the lepton mass eigenstate
basis (i.e.\ the muon $A$-term).  The last diagram employs flavor
violation at $\mu_L$ and $\mu_R$ vertices and is proportional to
$V_{\mu_R \tilde{\tau}_R}\tilde{A}_{\tilde{\tau}_R \tilde{\tau}_L}
V_{\tilde{\tau}_L \mu_L}^\dagger$, where $\tilde{A}_{\tilde{\tau}_R
\tilde{\tau}_L}$ is the 33 element of the $A$-term matrix, written in
the mass eigenstate basis of {\bf s}leptons, and $V\sim UT^\dagger$
are the lepton/slepton mixings (precisely defined in appendix~A).  The
intermediate diagrams are proportional to $V_{\mu_R
\tilde{\tau}_R}\tilde{A}_{\tilde{\tau}_R \mu_L}$ and $\tilde{A}_{\mu_R
\tilde{\tau}_L} V_{\tilde{\tau}_L \mu_L}^\dagger$, that involve the
$A$-term matrix in the lepton-slepton mixed basis (their general
expressions in terms of the $U_L,U_R,T_L,T_R$ matrices can be found in
appendix A, eq.\eq{explicit}).  Similar expressions hold for $d_e$.
We now compute these flavor factors.

We begin with diagonalizing explicitly the charged lepton mass matrix.
The values of $\lambda_e,\lambda_\mu,\lambda_\tau>0$ are related to
the complex elements of the lepton Yukawa matrix $\lambda_{ij}$ by
\begin{equation}\label{eq:eigenvalues}
\lambda_{\tau} e^{i\phi_\tau}  = \lambda_{33} ,\qquad
\lambda_{\mu} e^{i\phi_\mu} = \lambda_{22}-\epsilon_{23}\epsilon_{32} \lambda_{\tau} e^{i\phi_\tau},\qquad
\lambda_{e} e^{i\phi_e}  = \lambda_{11}
-\epsilon_{12}\epsilon_{21} \lambda_{\mu} e^{i\phi_\mu}
\end{equation}
where
$$
\epsilon_{23} = \frac{\lambda_{23}}{\lambda_{33}} ,\qquad
\epsilon_{32} = \frac{\lambda_{32}}{\lambda_{33}} ,\qquad
\epsilon_{12} = \frac{\lambda_{12}}{\lambda_{\mu} e^{i\phi_\mu}},\qquad
\epsilon_{21} = \frac{\lambda_{21}}{\lambda_{\mu} e^{i\phi_\mu}} 
$$
are the complex $ij$-rotation angles in $U^R_{32}$, $U^L_{23}$,
$U^R_{21}$, $U^L_{12}$ respectively. Our equations hold if
$|\epsilon_{ij}|\ll 1$, otherwise they are still qualitatively
correct.  In the standard basis where $\lambda_E = {\rm
diag}(\lambda_e,\lambda_\mu,\lambda_\tau)$, the lepton $A$-term
matrix, that is the factor entering in the first contribution to $d_e,d_\mu$ (fig.\fig{dipoli}a), is
\begin{equation}
\label{eq:Aff}
U_R\A_E U^\dagger_L = 
\begin{pmatrix}
\lambda_e A_e &
\epsilon_{12}\lambda_\mu e^{i(\phi_\mu-\phi_e)} (A_{12}-A_\mu) &
-\epsilon_{12}\epsilon_{23} \lambda_\tau e^{i(\phi_\tau-\phi_e)}
(A_{23}-A_{33})\\ 
\epsilon_{21}\lambda_\mu (A_{21}-A_\mu) &
\lambda_\mu A_\mu &
\epsilon_{23} \lambda_\tau e^{i(\phi_\tau-\phi_\mu)} (A_{23}-A_{33}) \\
-\epsilon_{21}\epsilon_{32} \lambda_\tau (A_{32}-A_{33}) &
\epsilon_{32} \lambda_\tau (A_{32}-A_{33}) &
\lambda_\tau A_{33}
\end{pmatrix} \; ,
\end{equation}
where
\globallabel{eq:Amu}
\begin{align}
A_\mu & = A_{22} + a_\mu (A_{22}-A_{23}-A_{32}+A_{33}) ,
& a_\mu &\equiv \epsilon_{23}\epsilon_{32} \frac{m_\tau}{m_\mu} 
e^{i(\phi_\tau-\phi_\mu)},  \mytag \\
A_e  &= A_{11} + a_e (A_{11}-A_{21}-A_{12}+A_{\mu}),
& a_e &\equiv \epsilon_{12}\epsilon_{21} \frac{m_\mu}{m_e} 
e^{i(\phi_\mu-\phi_e)} . \mytag
\end{align}
The quantities $\tilde A_{\mu_R\mu_L}$, $\tilde A_{e_R e_L}$ entering
the first contribution to the EDMs (fig.~\ref{fig:dipoli}a) are given
respectively by $\lambda_\mu A_\mu$ and $\lambda_e A_e$. The
coefficients $a_e$ and $a_\mu$ are texture-dependent complex numbers,
that measure how significant are the off-diagonal contributions to the
$e$ and $\mu$ mass, respectively.  Their natural range is $0 <
|a_\ell| <\hbox{few}$.  In particular, one has $a_e = -1$ for a
texture with a vanishing 11 element.  The position of phases in
eq.\eq{Aff} reflects the choice of including the matrix of phases
$\Phi$ in the right-handed rotation.

The two order one contributions (fig.\fig{dipoli}b) need the elements of the $A$-term matrix in the mixed basis in which
one $\tilde{\tau}$ is involved:
\begin{equation}\label{eq:explicit2}
\begin{array}{ll}
\A_{\tilde\tau_R \mu_L} = 
\epsilon_{32} \lambda_\tau  (A_{32}-A_{33})e^{i\phi_\tau}, &
\A_{\tilde\tau_R e_L} = - \epsilon_{21}\A_{\tilde\tau_R \mu_L}, \\
\A_{\mu_R \tilde\tau_L} = 
\epsilon_{23} \lambda_\tau  (A_{23}-A_{33})e^{i(\phi_\tau-\phi_\mu)},&
\A_{e_R \tilde\tau_L} = - \epsilon_{12}\A_{\mu_R \tilde\tau_L}.
\end{array}\end{equation}
Finally, the order two contribution (fig.\fig{dipoli}c) involve the 33
element of the $A$-term matrix in the slepton mass basis, $\tilde
A_{\tilde\tau_R\tilde\tau_L}$, which essentially coincides with
$\lambda_\tau A_{33}$.

%

\medskip

From the above equations we can finally recover the expression for the
quantities in eq.~(\ref{dmucontr}) directly related to the EDMs.
For the muon EDM we get
\begin{eqnsystem}{sys:contrmu}
A_0^\mu   &=&  m_\mu A_{\mu} \\
A_{1L}^\mu   &=&
e^{i(\phi_\tau-\phi_\mu)} \epsilon_{32}
(\epsilon_{23}-\tilde\epsilon_{23}) m_\tau (A_{33}-A_{32})\\
A_{1R}^\mu   &=&
e^{i(\phi_\tau-\phi_\mu)} \epsilon_{23}
(\epsilon_{32}-\tilde\epsilon_{32}) m_\tau (A_{33}-A_{23})\\
A_2^\mu   &=& e^{i(\phi_\tau-\phi_\mu)}
(\epsilon_{23}-\tilde\epsilon_{23})
(\epsilon_{32}-\tilde\epsilon_{32}) m_\tau (A_{33} + \mu\tan\beta) ,
\end{eqnsystem}
where $\tilde\epsilon_{32}$ ($\tilde\epsilon_{23}$) represent the
contribution of the 23 rotation that diagonalizes the left-handed
(right-handed) sfermion mass matrix.  For the electron EDM we get
\begin{equation}\label{eq:contre}
A_0^e =
m_e A_e,\quad
A_{1L}^e = A_{1L}^\mu a_e m_e/m_\mu,\quad
A_{1R}^e = A_{1R}^\mu a_e m_e/m_\mu,\quad
A_2^e = A_{2}^\mu a_e m_e/m_\mu.
\end{equation}
%
%

We can now discuss the size of the different contributions to the
EDMs.  If the $A$-terms are complex, the trivial term due to $A_e$ and
$A_\mu$ (contained in $A_0$) would give $|d_e/d_\mu| \sim m_e/m_\mu$
for any texture of the lepton Yukawa matrix 
A significant enhancement of
$d_\mu/d_e$ would be possible only as a result of an accidental
cancellation in the expression for $d_e$. The higher order
contributions to the EDMs, $A_i^{e,\mu}$ with $i=\{1L,1R,2\}$, satisfy
$\hbox{Im}\,(A^e_i/m_e)= \hbox{Im}\,(a_e A^\mu_i/m_\mu)$ and therefore
would give $|d_e/d_\mu| \approx |a_e| m_e/m_\mu$.  A sufficiently
large $d_\mu$ is obtained if $|a_e|\ll1$ (i.e.\ with an appropriate
texture for $\lambda_E$) and if the $A_{1L,R},A_2$ contributions are
the dominant ones.
The $A_2$ contribution dominates if $\tan\beta$ is large, or if
$m_{\tilde{e},\tilde{\mu}}\gg m_{\tilde{\tau}}$. On the contrary,
these contributions are suppressed if $|a_\mu|$ (and the $\tilde{\epsilon}_{23}$ rotations) are small.
If the large mixing angle
observed in atmospheric neutrino oscillations comes from the charged
lepton Yukawa matrix (i.e.\ if $|\epsilon_{23}|\sim 1$), we expect $a_\mu\sim
1$. In this case it is not possible to neglect 13
rotations~\cite{pippa} and one has a significant $\tau\to \mu\gamma$
rate.

Since strong bounds exist on the phase of some of the soft terms
(e.g.\ the $\mu$ term) for reasonable sparticle masses~\cite{dipoliSoft},
it is more appealing to consider the case of real (but non-universal) soft terms.  
CP violation is contained in the non diagonal Yukawa
matrices, that can be diagonalized by flavor rotations with complex
mixing angles $\epsilon_{ij}$, as in\eq{rotazioni}.  In general
$\lambda_E$ contains 4 independent phases, reduced to the 2 phases of
$a_e$ and $a_\mu$, under our assumption that the 13 and 31 elements of
the Yukawa matrix are negligible.  In the physical basis where
$\lambda_E = {\rm diag} (\lambda_e,\lambda_\mu, \lambda_\tau)$, the
slepton soft-terms are no longer real, giving rise to sizable EDMs as
explicitly shown by eq.s~(\ref{sys:contrmu},\ref{eq:contre}).  The
phases of $A_e$, $A_\mu$ are naturally small since $\hbox{Im}\,a_\ell \ll 1$ unless the two contributions to
$\lambda_e,\lambda_\mu$ in eq.\eq{eigenvalues} are comparable.  With real $A$-terms, the zeroth
order contributions $A_0$ gives
\begin{eqnsystem}{sys:ImA}
\hbox{Im}\,A_0^e &=& m_e (A_{33}-A_{23}-A_{32}+A_{22})\hbox{Im}\,a_e
a_\mu + m_e (A_{22}-A_{12}-A_{21}+A_{11})\hbox{Im}\, a_e \\ 
\hbox{Im}\,A_0^\mu &=&m_\mu (A_{33}-A_{23}-A_{32}+A_{22})\hbox{Im} \,a_\mu.
\end{eqnsystem}
The higher order contributions to $d_e$ are also proportional to
$a_e$, so that a $d_\mu/d_e$ larger than $m_\mu/m_e$ is always
obtained if $|a_e|\ll 1$ (i.e.\ if the electron mass is dominantly due
to the 11 element of $\lambda_E$).
A small $|a_e|$ could be due e.g.\ to a broken $L_e-L_\mu -L_\tau$ symmetry~\cite{nuTextures}
motivated by neutrino data (as recalled in section 3).
We remark, however, that $|a_e|\ll 1$ corresponds to a non vanishing
$\lambda_{11}$ dominating the electron mass. In the quark sector,
the analogous situation would be incompatible with the measured value of
$V_{us}$. 
It is interesting to study also the opposite situation of an
approximate texture zero in the 11 element of $\lambda_E$, so that
$a_e = -1$: in this case one always has $d_\mu/m_\mu= -d_e/m_e$
(rather than $d_\mu/m_\mu= +d_e/m_e$). Therefore, if the electron EDM
and supersymmetry were discovered, a measurement of $d_\mu/d_e$ (and in
particular of its sign) would be interesting, since it contains informations on the origin of the EDMs and,
eventually, on the structure of the lepton Yukawa matrix.  In the
quark sector the analogous measurement of $d_s/d_d$ could be performed
if hadronic uncertainties~\cite{PR} could be kept sufficiently under
control and if $d_u$ and $\theta_{\rm QCD}$ are small enough.

\medskip

We can finally pass to numerical results.
How large can be $d_\mu$,
compatibly with the most recent bounds on related processes collected in table 1?
It is convenient to translate the bounds on rare $\tau$ and $\mu$ decays
into bounds on the transition dipoles $d_{\ell\ell'}$, defined by the effective Lagrangian
\begin{equation}\label{eq:AmuSU5}
\mathscr{L}_{\rm eff}=
\frac{1}{2}\sum_{\ell\ell'}\big[\bar{\ell}_R
\gamma_{\mu\nu}d_{\ell\ell'} \ell'_L + 
\bar{\ell}'_L \gamma_{\mu\nu} d_{\ell\ell'}^*  \ell_R \big]F^{\mu\nu}
\end{equation}
where $\ell,\ell'=\{e,\mu,\tau\}$ and $\ell_{R(L)} = \frac{1}{2}(1\pm
\gamma_5) \ell$. When $m_\ell >m_{\ell'}$ the dipoles induce
the decay $\Gamma(\ell\to \ell' \gamma) = m_{\ell}^3(|d_{\ell\ell'}|^2
+ |d_{\ell'\ell}|^2)/(16\pi)$.  For $\ell = \ell'$
they induce the electric dipole $d_{\ell} = \hbox{Im}\, d_{\ell\ell}$
and the magnetic moment $ a_{\ell} = (2m_\ell/e) \hbox{Re}\,d_{\ell\ell}$.
Assuming equal left and right-handed mixing one has
$$|d_{e\mu}| = |d_{\mu e} |  < 2~10^{-26}\ecm,\qquad
|d_{\tau\mu}| = |d_{\mu \tau} |  < 5~10^{-22}\ecm.$$
In view of the many unknown parameters
(the sparticle masses, their mixings,\ldots) at the moment
we do not find useful to present detailed numerical results.
The dipoles are roughly given by
$$d_\ell \sim  \frac{e \alpha_{\rm em}m_\ell A_\ell}{24\pi\cos^2\theta_{\rm W}m_{\tilde{\ell}}^3}
\hbox{Im}\, a_\ell,\qquad
d_{\mu\tau}\sim d_{\tau\mu} \sim {d_\mu}\sqrt{\frac{m_\tau}{a_\mu m_\mu}},\qquad
d_{\mu e}\sim d_{e \mu} \sim {d_\mu}\sqrt{\frac{a_e m_e}{m_\mu}}
$$ 
where $A_\ell$ represent the combinations of $\mu$ and $A$ terms computed above (eventually enhanced by
$\tan\beta$) and $a_e, a_\mu$ are the most relevant texture-dependent free parameters.
The numerical factor is the one appropriate for the bino contribution
to the zeroth-order term $A^\ell_0$, evaluated for $m_{\tilde{\ell}} = m_{\tilde{B}}$.
For $A_\mu = m_{\tilde{\mu}} = 100\GeV$ and $a_\mu \sim 1$ one has $d_\mu \approx 0.3~10^{-22}\ecm$.
A $d_\mu$ in the $10^{-22}\ecm$ range is easily obtained and is compatible
with bounds on $\tau\to \mu \gamma$ for reasonable values of
$\epsilon_{23},\epsilon_{32},\tilde{\epsilon}_{23},\tilde{\epsilon}_{32}$
(such that $a_\mu \sim 1$)
and with bounds on $\mu\to e \gamma$ and $d_e$ for an appropriate texture of $\lambda_E$
with small 12 and 21 entries
(such that $|a_e|\ll 1$).
A $d_\mu$ in the $10^{-21}\ecm$ range requires in addition a moderately large $\tan\beta$ and
accidental cancellations in $d_{\mu\tau}, d_{\tau\mu}$.

\section{Conclusions}\label{conclusions}
Motivated by the prospects of improving the experimental sensitivity to the muon EDM by several orders of magnitude,
we have discussed the expectations for the muon and electron EDMs in supersymmetric scenarios.
If the EDMs scale with masses, $d_\mu/d_e = m_\mu/m_e$, the present limit on $d_e$ implies
$d_\mu \circa{<} 0.3~10^{-24}\ecm$.
We studied if a larger muon EDM, $d_\mu/m_\mu \gg d_e/m_e$, can be obtained.
The
generic answer is of
course yes: the MSSM has $\sim 100$ free parameters and one
just needs to assume that the appropriate ones (e.g.\ $A_\mu$, the
muon $A$-term) have a large complex phase.  However this possibility is not appealing.  
It is well known that the SUSY-breaking soft terms must satisfy
some highly constrained structure
in order to give an
acceptable phenomenology.
When
a scenario able to account for these constraints is considered, it is
not obvious that large effects and deviation from na\"{\i}ve scaling can be
obtained. 
For example, if the soft
terms are universal one has $A_\mu \approx A_e$ and $m_{\tilde{e}}\approx
m_{\tilde{\mu}}$ so that the relation $d_\mu/m_\mu \approx + d_e/m_e$ cannot be evaded.
It is therefore important to understand if a detectable $d_\mu$ can be obtained in frameworks able to account for
the tight constraints on the structure of supersymmetry breaking,
like some appropriate flavor symmetry or a flavor blind mechanism of
SUSY breaking.

The first case can be realized with an appropriate non-abelian flavor symmetry.
If the soft terms are complex the typical prediction for the EDM ratio is $d_\mu/m_\mu \sim
d_e/m_e$, so that $d_\mu/d_e$ could only be enhanced by accidental
order one factors with respect to the na\"{\i}ve value $m_\mu/m_e$.  In
particular, one gets $d_\mu/m_\mu =- d_e/m_e$ if the electron mass
arises from the 12 or 21 elements of the lepton Yukawa matrix.  If
instead the lepton Yukawa matrix has negligible 12 or 21 entries, some
contributions to $d_e$ and $d_\mu$ would naturally give
$|d_\mu/d_e|\gg m_\mu/m_e$, if they were dominant,
as it happens in some cases (e.g.\ large $\tan\beta$ or
$m_{\tilde{e},\tilde{\mu}}\gg m_{\tilde{\tau}}$).
If the soft terms are real (or almost real, as suggested by bounds on EDMs)
but have a non universal structure  (like the one allowed by flavor symmetries), 
sizable EDMs are obtained from the phases of
the Yukawa matrices.  The reason is that the complex flavor rotations
that diagonalize the lepton masses do not diagonalize, at the same
time, the slepton masses.
In this case, a lepton Yukawa matrix with small 12 or 21 entries
naturally enhances $d_\mu/d_e$ by suppressing $d_e$.
Despite this suppression, $d_e$ can saturate its experimental bound,
and a $d_\mu$ as large as $10^{-22}\ecm$ can be obtained
compatibly with all other bounds.
Unacceptably large effects are avoided because the induced phases
are typically small.

In the case of a flavor-blind mechanism of supersymmetry breaking,
CP-violation can be imprinted in the initially universal soft terms by
radiative corrections due to some higher energy physics.  The top
quark Yukawa coupling in SO(10) generates a sizable $\mu\to e \gamma$
decay rate and a sizable $d_e$: it also gives rise to $|d_\mu/d_e|\sim
|V_{td}/V_{ts}|^2$ times unknown ${\cal O}(1/3 \div 3)$ Clebsh-Gordon
factors.  Only with favorable factors it gives a $d_\mu$ slightly
above the planned sensitivity.  
If the lepton Yukawa matrix has 11, 13 and 31 texture zeros, these factors
always combine to give $d_\mu/d_e  = -m_\mu/m_e$.
In the see-saw context, radiative
effects are generated by the neutrino Yukawa couplings. 
Since they are uncertain, it is not impossible to obtain a sizable $d_\mu$.
However, the minimal see-saw structure suggested by neutrino data
(with the additional assumption that one unknown Yukawa coupling is
large enough) generates significant lepton-flavour violating effects in $\mu$ and $\tau$, but
without CP violation.  Large CP-violating effects in $\mu$ and $\tau$
are instead naturally obtained in a minimal SU(5)-unified see-saw, where a
$d_\mu$ up to about $10^{-23}\,e\,{\rm cm}$ is naturally obtained
compatibly with all other bounds.

In both cases, the maximal value of $d_\mu$ is typically accompanied by a 
$\tau\to \mu \gamma$ rate close to its experimental bound.
Finally, we also updated predictions for $\mu\to e \gamma$ rates and related processes
in SU(5) and SO(10) models,
and critically re-examinated which `predictions' are possible in the see-saw context.

\paragraph{Acknowledgements}
We thank J.\ Hisano for a comparison 
of the results of the numerical code for $\mu\to e \gamma$
and C. Savoy for useful comments.
A.R. would like to thank the
Aspen center for physics, where part of this work was done, for its
hospitality.

\appendix

\section{EDMs of charged leptons at one loop}

We are interested in EDMs generated by complex phases in fermion and sfermion mass matrices.
We therefore assume that the gaugino masses, the $\mu$ and $B\mu$ terms are real.
%
Only neutralino exchange contributes to EDMs of
charged leptons at one loop, giving
 \begin{equation}
 \label{eq:dmugeneral} 
  d_\ell =   \sum_{n,I}\frac{e\alpha_{\text{em}}}{4\pi \cos^2\theta_{\rm W}  M_{N_n}}
  \im\left[\cu^\pdagger_{\ell_L E_I} \cu^\dagger_{E_I\ell_R} \right]
  H_{n\tilde{B}} (H_{n\tilde{B}}+\cot\theta_{\rm W} H_{n\tilde{W}_3}) 
  \gfun\left(\frac{m^2_{E_I}}{M^2_{N_n}} \right),
\end{equation}
where $g(r)\equiv [1-r^2+2 r\ln r]/2/(1-r)^3$, $n=\{1,\ldots,4\}$,
$H_{n\tilde{B}}$, $H_{n\tilde{W}_3}$ are elements of the $4\times4$ matrix $H$ diagonalizing the
neutralino mass matrix, $H^T M_N H = \hbox{diag}(M_{N_1}\ldots
M_{N_4})$, $\cu$ is the $6\times 6$ unitary matrix diagonalizing the
charged slepton mass matrix $M^2_E$
(written in the supersymmetric basis where 
the mass matrix of charged leptons is diagonal)
as $\cu^\dagger M^2_E \cu =
\hbox{diag}(m^2_{E_1}\ldots m^2_{E_6})$.

In order to obtain a more useful expression, we diagonalize the
$6\times 6$ slepton mass matrix by treating the $A$-terms and the lepton masses 
as perturbations.  
In the approximation in which the first two slepton
families are degenerate, we then find for the dipoles $d_\ell$,
$\ell=e, \mu$, 
\begin{eqnarray}\label{dmufirstorder}
  d_\ell &=& - \,\frac{e\alpha_{\text{em}}}{4\pi \cos^2\theta_{\rm W}} 
\im\left[
A_2^\ell\,  G (\tilde\tau_R-\tilde\ell_R,
{\tilde\tau_L}-{\tilde\ell_L}) +\right.\\
&&+\left.
A_{1L}^\ell\,  G ({\tilde\tau_R}-{\tilde\ell_R},
{\tilde\ell_L}) + 
A_{1R}^\ell\,  G ({\tilde\ell_R},
{\tilde\tau_L}-{\tilde\ell_L}) + 
A_0^\ell\,  G ({\tilde\ell_R}, {\tilde\ell_L}) \right],   
\nonumber
\end{eqnarray}
where 
$$
G(a,b) \equiv \sum_{n=1}^4 \frac{H_{n\tilde{B}}}{M^3_{N_n}} (H_{n\tilde{B}}+\cot\theta_{\rm W} H_{n\tilde{W}_3}) 
g\big({m^2_a\over M^2_{N_n}} ,{m^2_b\over M^2_{N_n}}\big),\qquad
g(a,b) \equiv \frac{g(a)-g(b)}{a-b}
$$
$$G(a-a',b)\equiv G(a,b)-G(a',b),\qquad
G(a,b-b')\equiv G(a,b)-G(a,b'),$$ 
\begin{equation}\label{dmucontr}
\begin{array}{ll}
A_{1R}^\ell   = v \cos\beta \A_{\ell_R \tilde\tau_L}V^\dagger_{\tilde\tau_L\ell_L},\qquad&
A_0^\ell      = v \cos\beta \A_{\ell_R \ell_L},  \\
A_{1L}^\ell   = v \cos\beta \, V_{\ell_R\tilde\tau_R}	\A_{\tilde\tau_R\ell_L},&
A_2^\ell      = v V_{\ell_R\tilde\tau_R} (
	\cos\beta \A_{\tilde\tau_R \tilde\tau_L}
	+\mu\, \sin\beta\, 
	\lambda_{\tilde\tau_R \tilde\tau_L}  )
	V^\dagger_{\tilde\tau_L\ell_L}  \; .
\end{array}
\end{equation}
Here, $\lambda$ is the charged lepton Yukawa matrix and $\A$ are the
corresponding trilinear soft terms. The flavor basis in which they are
written is identified by their indexes ($\ell_{L,R}$, $\tau_{L,R}$
denote the lepton mass eigenstates, whereas $\tilde\ell_{L,R}$,
$\tilde\tau_{L,R}$ denote the slepton mass eigenstates). The matrices
$V$ measure the lepton-slepton mixing. In terms of the unitary
matrices $U_L,U_R,T_L,T_R$ defined in eq.\eq{rotazioni}, we have
%
%
%
\begin{equation}\label{eq:explicit}
 \begin{array}{lll}
\A_{\ell_R \ell_L}  = (U_R\A_E U^\dagger_L)_{\ell_R \ell_L},\quad&
\A_{\tilde\tau_R \ell_L} = (T_R \A_E U^\dagger_L)_{\tilde\tau_R \ell_L},\quad&
\A_{\ell_R \tilde\tau_L} = (U_R \A_E T^\dagger_L)_{\ell_R \tilde\tau_L} \\
\A_{\tilde\tau_R\tilde\tau_L} = (\A_E)_{\tilde\tau_R\tilde\tau_L},&
V_{\ell_L \tilde\tau_L} = ( U_L T^\dagger_L )_{\ell_L \tilde\tau_L},&
V_{\ell_R \tilde\tau_R} = ( U_R T^\dagger_R )_{\ell_R \tilde\tau_R}
\end{array}
\end{equation}
If
$\tan\beta$ is so large that $m_\tau\tan\beta \sim m_{\tilde{\tau}}$
one needs a more complicated expression, but there is no qualitative
change.

\section{One loop  effects in SO(10) and SU(5) unified models}\label{GUTs}

In SO(10) models, the unified top quark Yukawa coupling induces
lepton flavor violations in left-handed and right-handed slepton mass matrices.
Eq.~(\ref{dmufirstorder}) can be further
simplified taking into account the peculiar structure of the
flavor-violating $A$-terms generated by SO(10) effects (summarized in
eq.~(28) of~\cite{LFV}).  The relevant penguin dipoles $d_{\ell\ell'}$ (defined as in eq.\eq{AmuSU5})
are the ones that involve the first two generations, given by
$d_{\ell\ell'} = V_{\ell_R\tilde{\tau}_R} V_{\ell'_L \tilde{\tau}_L}
V_{\tau_R \tilde{\tau}_R}^* V_{\tau_L \tilde{\tau}_L}^* F$ where
\begin{eqnarray}
F &=& \displaystyle\frac{\alpha_{\rm em}m_\tau }{4\pi\cos^2\theta_{\rm W}}\nonumber
\bigg\{
(A_e+\mu\tan\beta)     G(\tilde{e}_L,\tilde{e}_R)+
(A_\tau + \mu\tan\beta)G(\tilde{\tau}_L,\tilde{\tau}_R)+\\&& -
(\frac{A_e + A_\tau}{2} + \mu\tan\beta)[G(\tilde{\tau}_L,\tilde{e}_R)+G(\tilde{e}_L,\tilde{\tau}_R)]\bigg\}.
\end{eqnarray}
In SU(5) models, the unified top quark Yukawa coupling induces
lepton flavor violations in the right-handed slepton mass matrix.
The transition dipoles $d_{\ell\ell'}$ can give rise to detectable
$\mu\to e \gamma$ and $\tau\to\mu\gamma$ rates, but
no sizable lepton EDMs are induced.
The transition dipoles that can induce significant effects are
$d_{e\mu} = V_{\mu_R \tilde{\tau}_R}^*  V_{e_R \tilde{\tau}_R} F$ and
$d_{\mu\tau} = V_{\tau_R \tilde{\tau}_R}^*V_{\mu_R \tilde{\tau}_R} F$
where now
\begin{eqnarray}\nonumber
F &=& \frac{\alpha_{\rm em}m_\mu }{4\pi\cos^2\theta_{\rm W}}\sum_{n=1}^4
\bigg\{ \frac{H_{n\tilde{B}}^2}{M_{N_n}^2}
\bigg[f(\frac{m_{\tilde{\tau}_R}^2}{M_{N_n}^2})-f(\frac{m_{\tilde{e}_R}^2 }{M_{N_n}^2})\bigg]+
\frac{H_{n\tilde{B}}}{M_{N_n}^3}
(H_{n\tilde{B}}+\cot\theta_{\rm W}H_{n\tilde{W}_3})\times
\\ &&\times \bigg[
(A_e+\mu\tan\beta)     
g(\frac{m_{\tilde{e}_L}^2}{M_{N_n}^2},  \frac{m_{\tilde{e}_R}^2}{M_{N_n}^2})-
(A_\tau + \mu\tan\beta)
g(\frac{m_{\tilde{e}_L}^2}{M_{N_n}^2},   \frac{m_{\tilde{\tau}_R}^2}{M_{N_n}^2})\bigg]+
\\ && -\frac{H_{n\tilde{B}}H_{n\tilde{h}_{\rm d}}}{M_{N_n}M_Z\cos\beta\sin\theta_{\rm W}}
\bigg[g(\frac{m_{\tilde{\tau}_R}^2}{M_{N_n}^2})-g(\frac{m_{\tilde{e}_R}^2}{M_{N_n}^2})\bigg]\bigg\}.
\nonumber
\end{eqnarray}
and  $f(r)=-[2+3r-6r^2+r^3+6r\ln r]/[6(r-1)^4]$.



\footnotesize
\begin{multicols}{2}

\end{multicols}

\end{document}